\documentclass[aps,prb,twocolumn,superscriptaddress,nofootinbib,longbibliography]{revtex4-1}
\usepackage[T1]{fontenc}
\usepackage[latin9]{inputenc}
\usepackage{array}
\usepackage{multirow}
\usepackage{amsmath}
\usepackage{amssymb}
\usepackage{graphicx}

\makeatletter
\pdfpageheight\paperheight
\pdfpagewidth\paperwidth

\providecommand{\tabularnewline}{\\}

\makeatother

\usepackage{amsmath,amsfonts,amssymb,color}
\usepackage{amsthm}
\usepackage{leftidx}
\usepackage{graphicx}
\usepackage{xcolor}
\usepackage{dcolumn}
\usepackage{bm}
\usepackage{epstopdf}
\usepackage{epsfig}
\usepackage{environ}
\usepackage{pdfcomment}

\usepackage{float}
\usepackage[T1]{fontenc}
\usepackage[latin9]{inputenc}
\usepackage{setspace}
\usepackage{esint}

\hypersetup{colorlinks=true,citecolor=blue,
linkcolor=blue,urlcolor=blue,pdfstartview=FitH,
bookmarksopen=true}

\begin{document}
\preprint{APS/123-QED}
\title{Non-Hermitian pseudo mobility edge in a coupled chain system}

\author{Sen Mu}
\email{senmu@u.nus.edu}
\affiliation{Department of Physics, National University of Singapore, 2 Science Drive 3, Singapore 117542, Singapore}

\author{Longwen Zhou}
\email{zhoulw13@u.nus.edu}
\affiliation{College of Physics and Optoelectronic Engineering, Ocean University of China, Qingdao, China 266100}

\author{Linhu Li}
\email{lilh56@mail.sysu.edu.cn}
\affiliation{Guangdong Provincial Key Laboratory of Quantum Metrology and Sensing $\&$ School of Physics and Astronomy, Sun Yat-Sen University (Zhuhai Campus), Zhuhai 519082, China}

\author{Jiangbin Gong}
\email{phygj@nus.edu.sg}
\affiliation{Department of Physics, National University of Singapore, 2 Science Drive 3, Singapore 117542, Singapore}

\date{\today}

\begin{abstract}
In this work, we explore interesting consequences arising from the coupling between a clean non-Hermitian chain with skin localization and a delocalized chain of the same length under various boundary conditions (BCs). We reveal that in the ladder with weak rung coupling, the non-Hermitian skin localization could induce a pseudo mobility edge in the complex energy plane, which separates states with extended and localized profiles yet allowing unidirectional transport of signals.  We also demonstrate the gradual takeover of the non-Hermitian skin effect in the entire system with the increase of the rung coupling under conventional open BC. When taking open BC for the non-Hermitian chain and periodic BC for the other, it is discovered that a quantized winding number defined under periodic BC could characterize the transition from the pseudo mobility edge to the trivial extended phases, establishing a ``bulk-defect correspondence'' in our quasi-1D non-Hermitian system. This work hence unveils more subtle properties of non-Hermitian skin effects and sheds light on the topological nature of the non-Hermitian localized modes in the proximity to systems with dissimilar localization properties.

\end{abstract}

\maketitle
\section{Introduction\label{sec:Int}}

Any amounts of uncorrelated disorder shall localize all bulk states in large noninteracting systems of dimension $d<3$~\cite{anderson1958,anderson1979}. In higher dimensions, a mobility edge (ME) emerges as the energy separating localized and extended eigenstates~\cite{mott1975,Lee1985,mott1987,evers2008,kondov2011,semeghini2015}. Replacing the uncorrelated disorder by a quasiperiodic potential, e.g., in the Aubry-Andr\'e-Harper model~\cite{aubry1980,harper1955}, a quantum phase transition can take place in one dimension where all eigenstates become extended or localized when the potential strength varies~\cite{soukoulis1982,das1986}. In the presence of long-range hopping, slowly varying potential or lattice deformations, a mobility edge may also appear in low-dimensional systems~\cite{das1988,das1990,biddle2010,ganeshan2015,deng2019,wang2020}. 
The emergence of mobility edges in non-Hermitian disordered systems under periodic boundary conditions have been discussed~\cite{Hatano1996,Hatano1997}. Mobility edges in the complex energy plane have also been reported in non-Hermitian quasiperiodic systems recently~\cite{zeng2017,longhi2019,yanxia2020,liu2020,zeng2020,liu2021,longhi2021,Zhou_2021,Zhou_2021_2,ZhouDNHQC2022}.

Effective non-Hermitian descriptions for classical and quantum systems have attracted great attention. In theory, non-Hermitian models can host a plethora of exotic phenomena absent in their Hermitian counterparts~\cite{ganainy2018,ashida2020,bergholtz2021,lee2014,Nakagawa2018,mu2020,zhang2021,Zhou_NHFTI1,Zhou_DQPT1}, e.g., their exceptional topology have been discovered and classified into enriched symmetry tables~\cite{shen2018,gong2018,katabata2019,zhou2019,wojcik2020}. Experimentally, various non-Hermitian phases have been observed in a wide range of {classical and} quantum simulators like photonic, acoustic, electronic and ultracold atom systems~\cite{zeuner2015,zhu2018,li2019,wu2019,xiao2020,helbig2020}, yielding potential applications such as topological lasing and high-performance sensing~\cite{gal2018,miguel2018,sebastian2020,Hodaei2017,Chen2017,McDonald2020,Budich2020}. 

Among the rich features of non-Hermitian models reported so far, the non-Hermitian skin effect (NHSE) stands out as an intriguing localization mechanism~\cite{Martinez2018,lee2019,song2019,yi2020,Okuma2020,lee2019hybrid,li2020topo,zhang2020,li2020,Zhou_PRB2021}. It refers to the phenomenon where all eigenstates are localized at the edges when the system breaks translational symmetry. Because of possibly betraying the well-established band theory and the renowned principle of bulk-boundary correspondence (BBC), new theoretical formulations, such as the non-Bloch Brillouin zone, and refined definitions of topological invariants have been proposed and pushed forward for a resolution in the presence of NHSE~\cite{Xiong2018,Kunst2018,yao2018edge,yao2018chern,yokomizo2019,song2019real,Borgnia2020,yang2020,Zhou_PRB2021}. 

Mobility edges without disorder or quasiperiodicity are rarely explored~\cite{zhang2021me,dwiputra2021}, and we naturally ask whether there exists a mobility edge for systems exhibiting non-Hermitian skin localization? In this work, we propose to consider a clean non-Hermitian skin localized chain with nonreciprocal hoppings coupled to a delocalized chain of the same length. We reveal that the NHSE in the non-Hermitian chain can propagate to its proximity under certain boundary conditions and interchain couplings. Remarkably, a mobility edge seemingly emerges in the complex energy plane separating the localized eigenstates from the extended, though a suppression of transport in the localized regime analogous to Anderson localization does not occur and is replaced by directional amplification instead~\cite{McDonald2018,Wanjura2020,xue2021,Li2021response}. We hence dub this border between localization and delocalization as a non-Hermitian pseudo mobility edge and investigate its properties under various boundary conditions. Surprisingly, a quantized jump of a spectral winding number is found to characterize the transition from a phase with a pseudo mobility edge to extended phases under an unconventional boundary condition, establishing a ``bulk-defect correspondence''. 

The remaining part of the paper is organized as follows. We introduce our model in Sec.~\ref{sec:Mod} with the importance of boundary conditions for non-Hermitian skin localization highlighted. In Sec.~\ref{sec:Res}, we demonstrate that the proximity effect of NHSE could induce a pseudo mobility edge phase in the composite system with weak interchain coupling under different boundary conditions. Characterizations of the eigenstate spatial profiles are performed by calculating the inverse participation ratio (IPR) and its scaling with the system size. Furthermore, the fractal dimension (FD) and topological winding number are also employed to confirm the pseudo mobility edge phase and transitions to extended phases. Finally, we summarize our results and discuss potential future directions in Sec.~\ref{sec:Sum}.

\section{Model\label{sec:Mod}}

\begin{figure}
\begin{centering}
\includegraphics[scale=0.49]{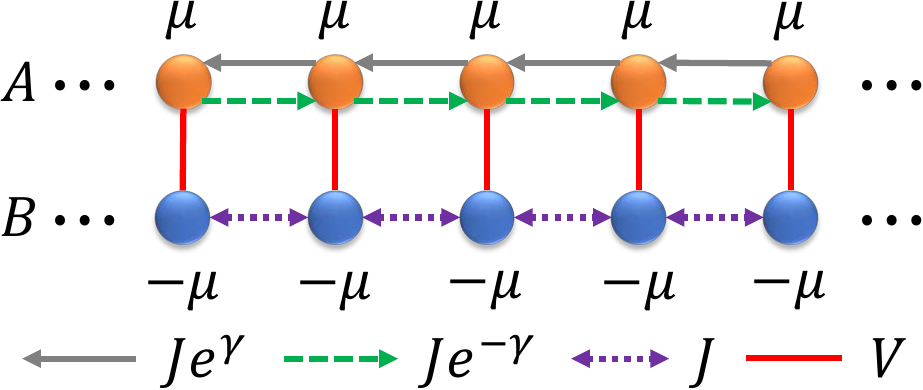}
\par\end{centering}
\caption{A schematic illustration of the hybridized HN ladder. The leg A/B
is formed by a clean HN/Hermitian chain.\label{fig:Model}}
\end{figure}

\begin{figure}
	\begin{centering}
		\includegraphics[width=0.85\linewidth]{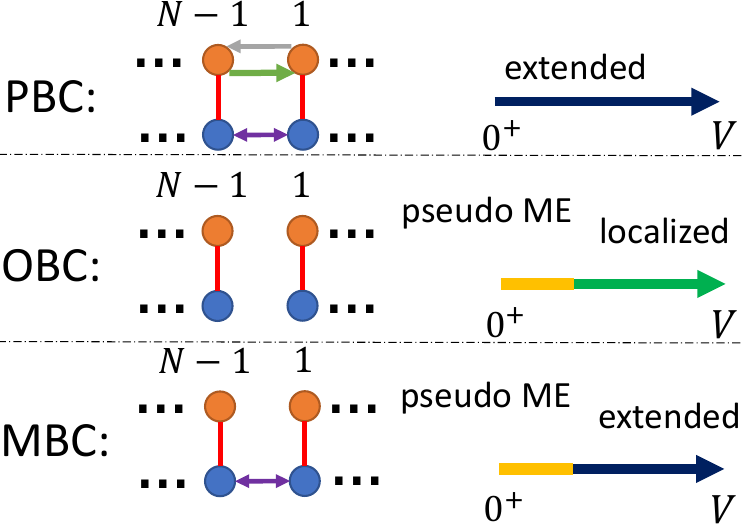}
		\par\end{centering}
	\caption{A sketch of possible eigenstate spatial profiles of the hybridized HN ladder under different BCs when the interchain coupling $V$ increases with other system parameters fixed. $N-1$ and $1$ are the indices of unit cells.\label{fig:sketch}}
\end{figure}

In this section, we introduce a minimal model that could demonstrate the presence of the pseudo mobility edge before the full takeover of NHSE. Our model consists of a non-Hermitian disorder-free Hatano-Nelson (HN) chain~\cite{Hatano1996,Hatano1997} with nonreciprocal nearest-neighbor hoppings and another chain with reciprocal hoppings of the same length. We further introduce site-to-site reciprocal single-particle hoppings between the two chains to realize a non-trivial system. We will term this model as a hybridized HN ladder thereafter, with a schematic illustration given in Fig.~\ref{fig:Model}. In the tight-binding representation, the bulk Hamiltonian of our hybridized HN ladder reads
\begin{alignat}{1}
H& = \sum_{n}J(e^{\gamma}c^\dagger_{n,A}c_{n+1,A}+e^{-\gamma}c^\dagger_{n+1,A}c_{n,A})\nonumber \\
& + \sum_{n}(Jc^\dagger_{n,B}c_{n+1,B}+Vc^\dagger_{n,A}c_{n,B}+{\rm H.c.})\nonumber \\
& + \sum_{n}\mu(c^\dagger_{n,A}c_{n,A}-c^\dagger_{n,B}c_{n,B}),\label{eq:H}
\end{alignat}
where $c^\dagger_{n,A/B}$ is the creation operator at $A$/$B$
sublattice in the $n$-th unit cell with $n=1,...,N$ the unit cell indices~(see Fig.~\ref{fig:Model}). We may alternatively think of the two sublattices as an additional degree of freedom such as spin or flavor. The parameter $J$ denotes the nearest-neighbor hopping amplitude and $\gamma$ controls the asymmetry between right-to-left and left-to-right hoppings along the HN chain formed by all $A$ sublattices. The interchain coupling strength is controlled by $V$, and an onsite potential bias $2\mu$ is applied between the two sublattices in each unit cell. From now on, we take $J=1$ as the unit of energy. Any additional terms from the boundary coupling is illustrated in Fig.~\ref{fig:sketch}.

Translational symmetry in our model can be either preserved by connecting the first and last sites, or broken by not connecting them, though it is always broken in Anderson models with disorder induced localization. Boundary conditions are crucial or a priori for the appearance of the non-Hermitian skin localization, and in this work, we will consider various types of boundary conditions for our system. 

We first note that under periodic boundary conditions~(PBCs), i.e. $c^\dagger_{N,s}=c^\dagger_{1,s}$ with $s=A,B$, our model is translational invariant and thus can be written as $H=\sum_{k}\tilde{H}(k)c^\dagger_kc_k$ in momentum space with the Fourier transformation $c^\dagger_{n,s}=\frac{1}{\sqrt{N}}\sum_{k}e^{-ikn}c^\dagger_{k,s}$ with $s=A,B$. We have
\begin{alignat}{1}
\tilde{H}(k)& = h_{0}\sigma_{0}+h_{x}\sigma_{x}+h_{z}\sigma_{z},\label{eq:Hk}\\
h_{0}& = \cos(k-i\gamma)+\cos k,\nonumber \\
h_{x}& = V,\nonumber \\
h_{z}& = \cos(k-i\gamma)-\cos k+\mu,\nonumber 
\end{alignat}
where the identity $\sigma_0$ and Pauli $\sigma_{x,z}$ matrices acting in the pseudo-spin-$\frac{1}{2}$ space spanned by sublattices $A$ and $B$ in each unit cell. $k\in(-\pi,\pi]$ is the quasimomentum. Eigenstates under the PBC are Bloch states and extended in real space.

The open boundary condition~(OBC) of our model refers to taking OBCs for both chains, disconnecting all the end sites. It is thus possible that all eigenstates are pushed by the nonreciprocal pumping in a preferred direction and accumulate at the boundaries. Here, we further consider taking PBC along the Hermitian chain and OBC along the HN chain, named as mixed boundary condition~(MBC). Our motivation for considering the MBC is that the chain with reciprocal hopping is expected to be not much affected by the boundary conditions, where eigenstates of it are all extended when there is no interchain coupling, either as standing waves under OBC or Bloch waves under PBC. However, the eigensystem under MBC might change dramatically compared with that under OBC due to the extreme sensitivity to extra coupling of the resultant non-Hermitian ladders. Intrinsic connections between results obtained under different types of boundary conditions will be presented and discussed in the following sections.

\section{Results and Discussions\label{sec:Res}}

We now present the eigensystem and the emergence of a pseudo mobility edge in the complex energy plane of our model under different boundary conditions. In particular, we will focus our attention on the topological features of the hybridized HN ladder under the unconventional MBC. 

In Sec.~\ref{subsec:E_pbc}, we show the spectrum of the system under PBC. Unique to non-Hermitian systems, a spectral winding number can be introduced to characterize the transition from a point-gap to a line-gap spectrum, with the phase boundary obtained analytically. In Sec.~\ref{subsec:E_obc_mbc}, we present in detail the eigensystem of our model under OBC and MBC. One of our main results is that a pseudo mobility edge phase, determined from the inverse participation ratio (IPR) of eigenstates, emerges when the interchain coupling $V$ is small, while it transits to a localized phase and extended phase as $V$ increases under OBC and MBC, respectively. We provide a sketch of phases based on eigenstate spatial profiles of our model in Fig.~\ref{fig:sketch} for the ease of reference. Bounds of the pseudo mobility edge under MBC are exactly determined by the edges of Bloch bands and further confirmed by FD calculations in Sec.~\ref{subsec:FD}. Especially, we find that the transition of the system from a pseudo mobility edge phase to an extended one under MBC is accompanied by the quantized jump of a spectral winding number under PBC. In Sec.~\ref{subsec:WN}, we establish the topological phase diagram of the model under MBC and discuss its transitions.

\subsection{Spectrum and winding number under PBC\label{subsec:E_pbc}}

\begin{figure}
\begin{centering}
\includegraphics[width=1\linewidth]{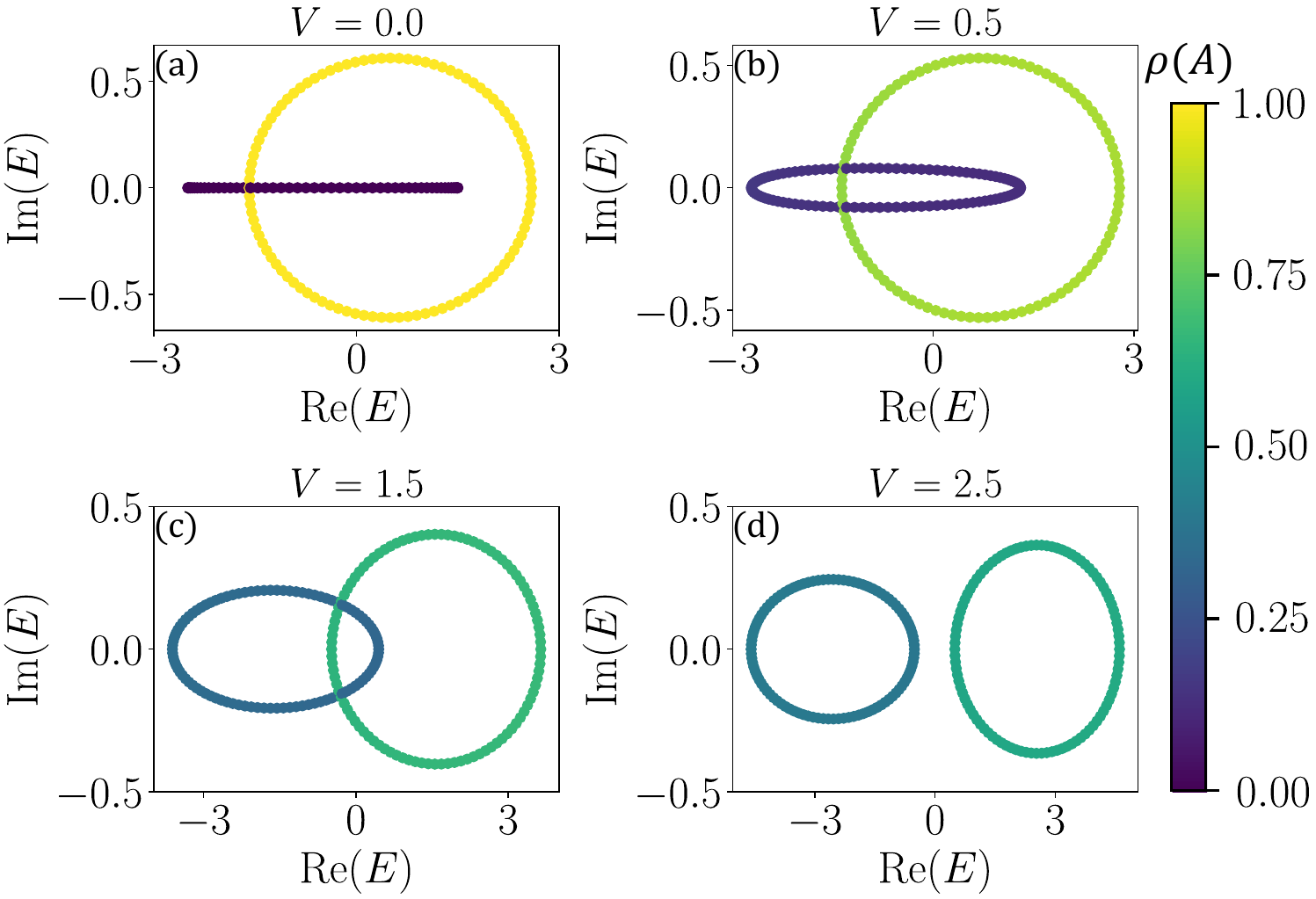}
\par\end{centering}
\caption{Spectrum of the hybridized HN ladder under PBC. Other system parameters are $(J,\gamma,\mu)=(1,0.3,0.5)$ for (a)--(d) and the length of ladder is $N=100$. The color of each point represents the occupation of the corresponding eigenstate on the HN chain.\label{fig:E_pbc}
}
\end{figure}

We first present the spectrum of our system under PBC in Fig.~\ref{fig:E_pbc}. From Eq.~(\ref{eq:Hk}), we see that the system has two complex bulk bands given by
\begin{equation}
E_{\pm}(k)=h_{0}\pm\sqrt{h_{x}^{2}+h_{z}^{2}}.\label{eq:Ek}
\end{equation}
Along the real-energy axis, the upper bounds of $E_{\pm}(k)$ are
located at
\begin{alignat}{1}
E_{+}(0)= & \cosh\gamma+1+\sqrt{V^{2}+(\cosh\gamma-1+\mu)^{2}},\label{eq:Ep0}\\
E_{-}(0)= & \cosh\gamma+1-\sqrt{V^{2}+(\cosh\gamma-1+\mu)^{2}},\label{eq:Em0}
\end{alignat}
whereas the lower bounds of $E_{\pm}(k)$ are found at
\begin{alignat}{1}
E_{+}(\pi)= & -\cosh\gamma-1+\sqrt{V^{2}+(\cosh\gamma-1-\mu)^{2}},\label{eq:Epp}\\
E_{-}(\pi)= & -\cosh\gamma-1-\sqrt{V^{2}+(\cosh\gamma-1-\mu)^{2}}.\label{eq:Emp}
\end{alignat}

When the interchain coupling is turned off, i.e., $V=0$, the spectrum of the HN chain forms a loop on the complex energy plane, and eigenenergies of the other chain reside along the real axis, as
shown in Fig.~\ref{fig:E_pbc}(a). When $V$ is turned on, the two chains are coupled and the spectrum of the hybridized system consists of two loops in the complex energy plane. We also show the occupation on the HN chain of each eigenstate $\rho(A)=\sum_{n=1}^{N}|\psi_{n,A}^{i}|^{2}$ by color to indicate how strong the two chains are hybridized. The two loops partially overlap, and they share the central point-gap in the range $E_{+}(\pi)<{\rm Re}(E)<E_{-}(0)$.
When $V$ is large, the two bands are shifted away and become separated by a line-gap, as shown in Fig.~\ref{fig:E_pbc}(d). The spectrum of the system changes from point- to line-gapped when $E_{-}(0)=E_{+}(\pi)$, yielding critical values of interchain coupling
\begin{equation}
V_{\pm}^{c}=\pm\frac{2}{\cosh\gamma+1}\sqrt{\cosh\gamma\left[(\cosh\gamma+1)^{2}-\mu^{2}\right]}.\label{eq:Vc}
\end{equation}
Note that we are more interested in the regime where the two chains can be efficiently coupled by even a small $V$, and hence we focus on the spectral transitions mentioned above with $\mu\in(\mu_{-},\mu_{+})$, where 
\begin{equation}
\mu_{\pm}=\pm(\cosh\gamma+1).\label{eq:MUpm}
\end{equation}
When $|\mu|>|\mu_{\pm}|$, the onsite bias is too large such that the bands are separated by a line-gap at $V=0$. Moreover, the band touching point at the spectral transition is found to be $E_{0}=E_{-}(0)=E_{+}(\pi)$, i.e.,
\begin{equation}
E_{0}=-\mu\frac{\cosh\gamma-1}{\cosh\gamma+1}\label{eq:EB}
\end{equation}
assuming $\mu\in(\mu_{-},\mu_{+})$. 

Now we can define a spectral winding number with respect to $E_{0}$ under PBC \cite{gong2018}, i.e.,
\begin{equation}
w=\int_{-\pi}^{\pi}\frac{dk}{2\pi i}\partial_{k}\ln\det\left[H(k)-E_{0}\right].\label{eq:WN}
\end{equation}
When $V\in(V_{-}^{c},V_{+}^{c})$, the base energy $E_{0}$ resides in the {central} 
point gap of $H(k)$ and is encircled twice by $E_{\pm}(k)$ when $k$ sweeps from $-\pi$ to $\pi$, yielding a winding number $w=\pm2$, which is easy to see geometrically in Figs.~\ref{fig:E_pbc}(b)--(c). When $|V|>|V_{\pm}^{c}|$, $E_{0}$ appears
in the line gap of $H(k)$ and is not encircled by $E_{\pm}(k)$, leading to a winding number $w=0$. Therefore, the point-gap and line-gap phases of the hybridized HN ladder under PBC can be distinguished by the topological winding number $w$. We will demonstrate a ``bulk-defect'' correspondence of this winding number in Secs.~\ref{subsec:FD} and \ref{subsec:WN}. Before that, we first resolve the spectral and transport nature of the system under the OBC and MBC.

\subsection{Eigensystem under OBC and MBC\label{subsec:E_obc_mbc}}

\begin{figure}
\begin{centering}
\includegraphics[width=1\linewidth]{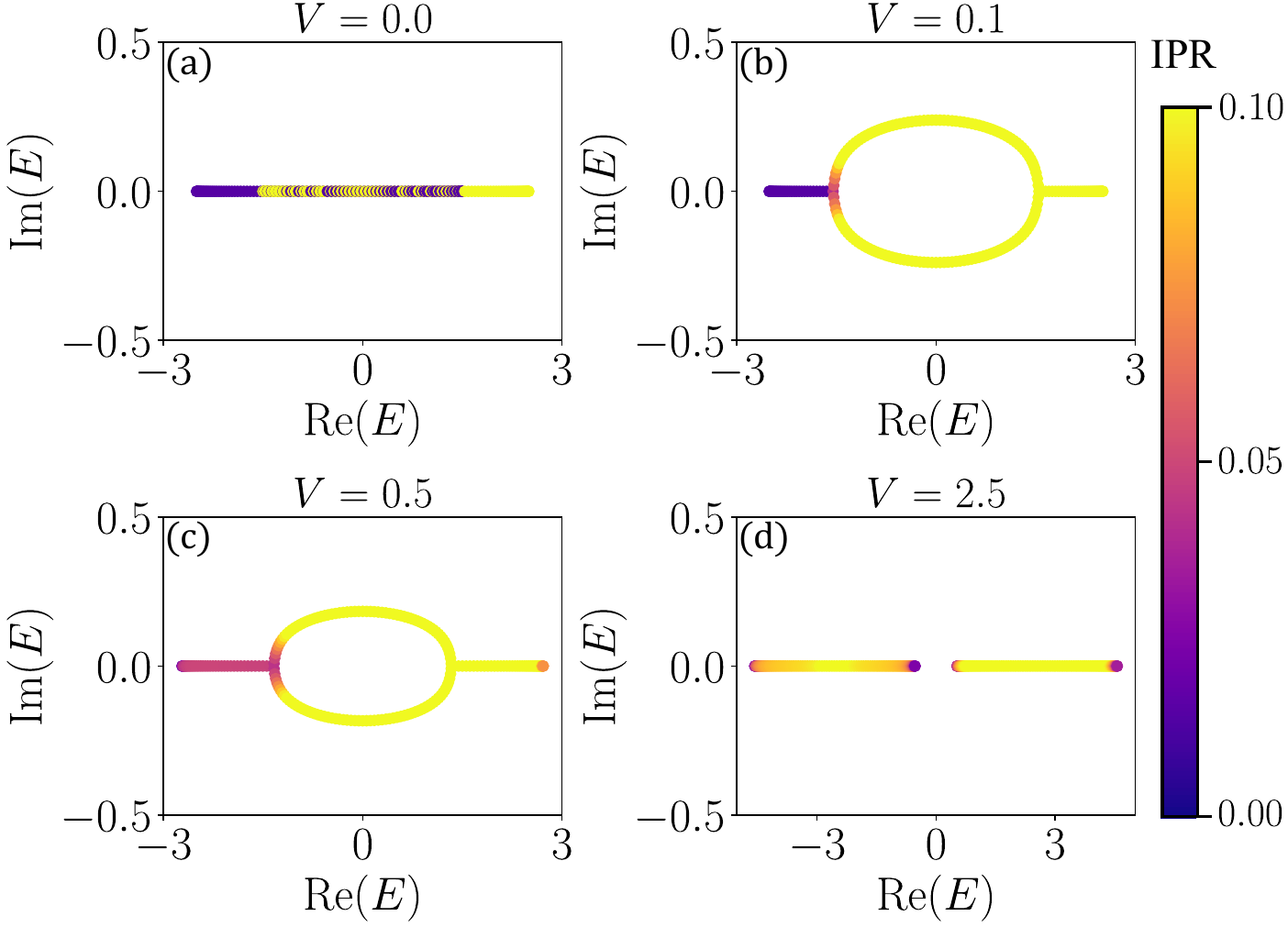}
\par\end{centering}
\caption{Spectrum of the hybridized HN ladder under OBC. Other system parameters are $(J,\gamma,\mu)=(1,0.3,0.5)$ for (a)--(d) and the length of ladder is $N=100$. The color of each point represents the IPR of the corresponding eigenstate. All states with ${\rm IPR}>0.1$ are dyed with the same color for illustration purpose.}
\label{fig:E_obc}
\end{figure}

\begin{figure}
\begin{centering}
\includegraphics[width=1\linewidth]{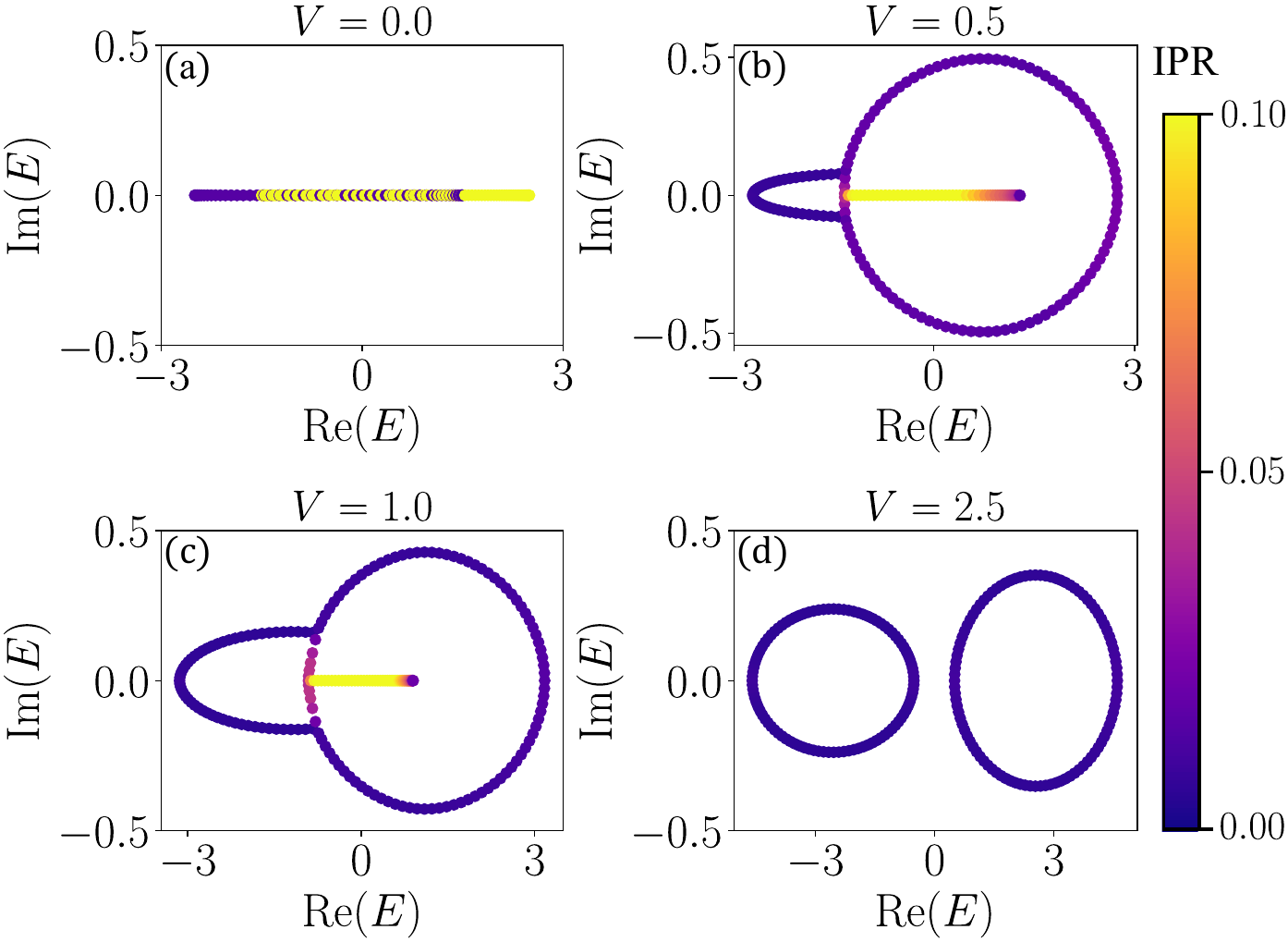}
\par\end{centering}
\caption{Spectrum of the hybridized HN ladder under MBC. Other system parameters are $(J,\gamma,\mu)=(1,0.3,0.5)$ for (a)--(d) and the length of ladder is $N=100$. The color of each point represents the IPR of the corresponding eigenstate. All states with ${\rm IPR}>0.1$ are dyed with the same color for illustration purpose.}
\label{fig:E_mbc}
\end{figure}

Under OBC or MBC, we can obtain the spectrum and eigenstates of the hybridized HN ladder by solving the eigenvalue equation with the corresponding Hamiltonian Eq.~(\ref{eq:H}) in real space. We present the spectrum for OBC in Fig.~\ref{fig:E_obc} and that for MBC in Fig.~\ref{fig:E_mbc}. We first note that when the two chains are decoupled, eigenenergies are all real since the spectrum of an open HN chain is real, as shown in Figs.~\ref{fig:E_obc}(a) and \ref{fig:E_mbc}(a). 

For a finite interchain coupling $V>0$, a dramatic change of the spectrum occurs, as a result of the extreme sensitivity of non-Hermitian systems. We observe a ring immediately develops in the central region of the OBC spectrum and shrinks as $V$ increases, as seen in Fig.~\ref{fig:E_obc}(b)-(c). Eventually, two line-gapped bands on the real axis are formed when $V$ is sufficiently large. 
We can understand such transitions qualitatively by examining the role of $V$ in different parameter regions. An infinitesimal coupling between two chains with dissimilar skin localization in thermodynamic limit could alter very much their eigenspectrum in a non-perturbative manner~\cite{li2020}. In principle, we can obtain the OBC spectrum analytically by employing the generalized Brillouin zone approach with a nonvanishing $V$ in thermodynamic limit, though it is highly nontrivial to derive the exact expression. Instead, on one hand, we could understand the emergence of the central ring qualitatively with the concept of level repulsion in standard quantum mechanics. In the overlapping region of the spectrum, density of states is high and nearby eigenvalues repel in the complex plane. Specifically, in the basis where the non-Hermitian Hamiltonian of two decoupled chains is diagonal, the interchain hoppings may consist of an anti-Hermitian part, forcing the eigenvalues to repel one another along the imaginary axis~\cite{Feinberg1997}. A similarity transformation to a Hermitian model is thus impossible. On the other hand, we could also understand it from a dynamical point of view. Let us imagine if we input a signal at one of the chains. The interchain couplings then provide propagation channels for the signal frequency that is on resonance with the other chain, possibly allowing an infinite directional amplification of the signal via forming shortcuts in the system. This dynamical instability is now manifested as complex eigenenergies in the central ring regime of the spectrum~\cite{McDonald2018,liang2021anomalous}.
In the limit of large $V$, the roles of the interchain coupling $V$ and the potential bias $\mu$ are effectively exchanged. The model is then dominated by the term $V\sigma_x$ in each unit cell, and the eigenstates become the symmetric and anti-symmetric combinations of the two sublattices $A$ and $B$, which are weakly perturbed by intrachain couplings. Each of these linear combinations naturally inherits the nonreciprocal hoppings from the HN chains, and also the OBCs of both chains. Therefore, the spectrum ends up residing on the real axis as if there are two copies of open HN chains.

We next look at the spectrum under MBC. Interestingly, besides the development of rings for small $V$, we find that a portion of the spectrum is pinned to the ${\rm Re}(E)$ axis, ending at the $E_{+}(\pi)$ and $E_{-}(0)$ of the PBC spectrum, which is quite different from the case under OBC. We further verify that the range of $V$ to observe these real eigenvalues coincides with the point-gap phase $V\in(V_{-}^{c},V_{+}^{c})$ under PBC. When $V>V_{+}^{c}$, all eigenvalues form two loops separated by a line gap, resembling the spectrum of two closed HN chains. 
As discussed previously, the system is now analogous of two HN chains formed by two linear combinations of $A$ and $B$, except that they also inherit the boundary couplings of the original Hermitian chain. Therefore, both of them have finite boundary couplings, leading to the PBC-like loops of the spectrum.

Now let us turn our attention to the localization property of the eigenstates at different energies under OBC and MBC. We recall that a non-Hermitian Hamiltonian $H$ with eigenvalues ${E_n}$ has a set of right eigenvectors $H|\psi_n\rangle=E_n|\psi_n\rangle$ as well as a set of left eigenvectors $\langle\phi_n|H=\langle\phi_n|E_n$. They together form a complete biorthonormal eigensystem. In this work, we focus on the non-Hermitian skin localization and we therefore employ only the right eigenstates of the Hamiltonian. Note that in disordered HN chains under PBC, delocalization has been investigated with both the left and right eigenvectors~\cite{Hatano1998}. We define the inverse participation ratio (IPR) of a right eigenvector as
\begin{equation}
{\rm IPR}(|\psi_{i}\rangle)=\sum_{n=1}^{N}\sum_{s=A,B}|\psi_{n,s}^{i}|^{4},
\end{equation}
where $|\psi_i\rangle=\sum_{n,s}\psi^{i}_{n,s}c^\dagger_{n,s}|0\rangle$ is the normalized right eigenvector of $H$. The IPRs of all eigenstates under OBC and MBC are indicated by the color of points in Figs.~\ref{fig:E_obc} and \ref{fig:E_mbc}, respectively.

\begin{figure}
\begin{centering}
\includegraphics[width=1\linewidth]{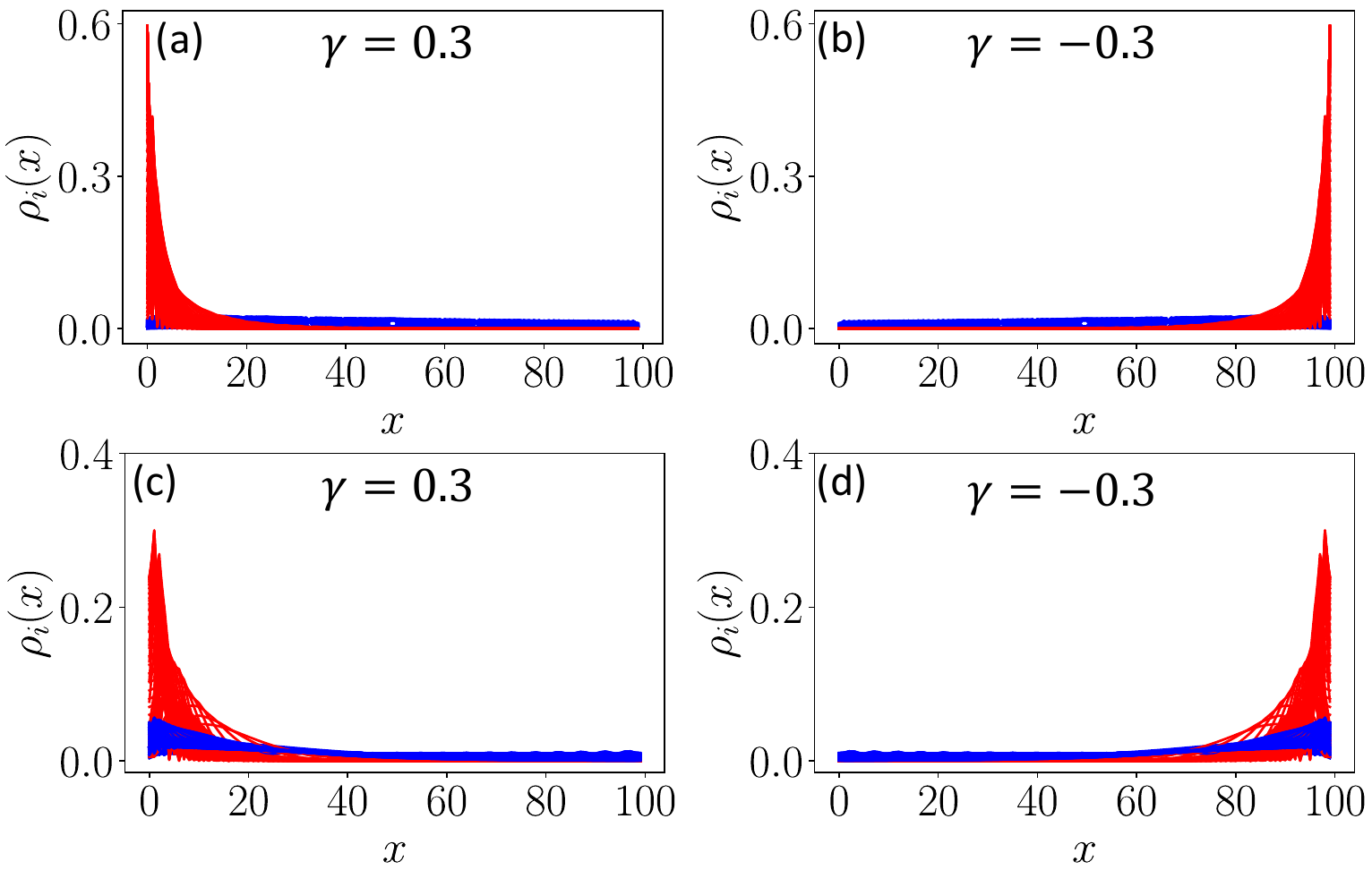}
\par\end{centering}
\caption{Eigenstate spatial distribution with different asymmetric parameter $\gamma$ under different boundary conditions. OBC for (a)-(b) with $V=0.1$ and MBC for (c)-(d) with $V=1.0$. Red colors are corresponding to eigenstates with ${\rm IPR}>0.02$ and blue colors to those with ${\rm IPR}<0.02$. Other system parameters are $(J,\mu)=(1,0.5)$ for (a)--(d) and the length of
ladder is $N=100$ with $L=2N$.
\label{fig:state_profiles}}
\end{figure}

One of the most striking features of the eigenstates of our model under OBC or MBC is the presence of both (skin) localized and extended states. This is evident from the sharp color contrast from their IPRs. A cluster of eigenstates under either OBC or MBC are more localized than others in a finite range of the interchain coupling $V$. In Fig.~\ref{fig:state_profiles}, we present the spatial distribution of right eigenstates defined as 
\begin{equation}
\rho_i(x)=\sum_{s=A,B}|\psi_{x,s}^{i}|^{2}
\end{equation}
for certain $V$ under both OBC and MBC. Note that the directional localization of some states towards one boundary for different asymmetric parameter $\gamma$ is a clear signature of NHSE, while the relation between a seemingly preferred direction indicated by $\gamma$ and the actual localization direction may not be trivial~\cite{li2022direction}. We also observe that the amount of localized eigenstates increases (decreases) when the interchain coupling is tuned up for OBC (MBC). Moreover, as these skin localized modes can coexist with extended states, we claim the NHSE-induced mobility edge (ME) on the complex energy plane for such coupled chain system. We note that there is a major difference between ME of non-Hermitian skin localization observed in our model under OBC and ME of Anderson localization in a disordered HN chain under PBC.  We do not expect the suppression of transport of local excitations in our system. Instead, with the presence of skin localized modes in the mobility edge phase, directional amplification is expected for signal frequencies matched with those energies~\cite{McDonald2018,Wanjura2020,xue2021,Li2021response}. This is the reasoning behind the term ``pseudo mobility edge'' we adopted here. Another observation is that when $V$ is large, all eigenstates are localized under OBC, while the localization of states eventually vanishes under MBC. This reconfirms  our previous interpretation at large $V$ in terms of effective HN chains without (with) boundary couplings for OBC (MBC). 

\begin{figure}
\begin{centering}
\includegraphics[width=1\linewidth]{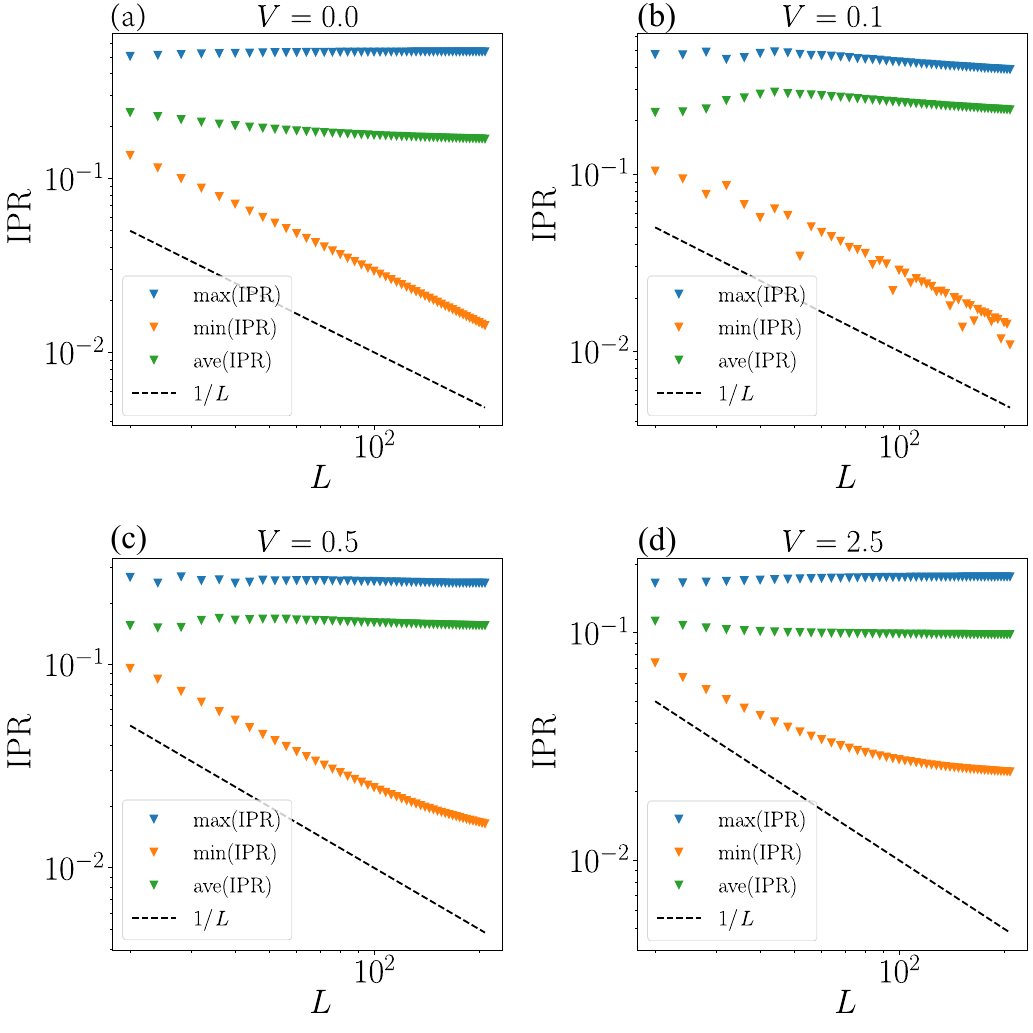}
\par\end{centering}
\caption{Scaling of the maximum, minimum and average of IPRs versus the length
of ladder $N$ under OBC. Other system parameters
are $(J,\gamma,\mu)=(1,0.3,0.5)$ for (a)--(d) and the length of
ladder goes from $N=10$ up to $N=100$ with $L=2N$.\label{fig:IPR_obc}}
\end{figure}

\begin{figure}
\begin{centering}
\includegraphics[width=1\linewidth]{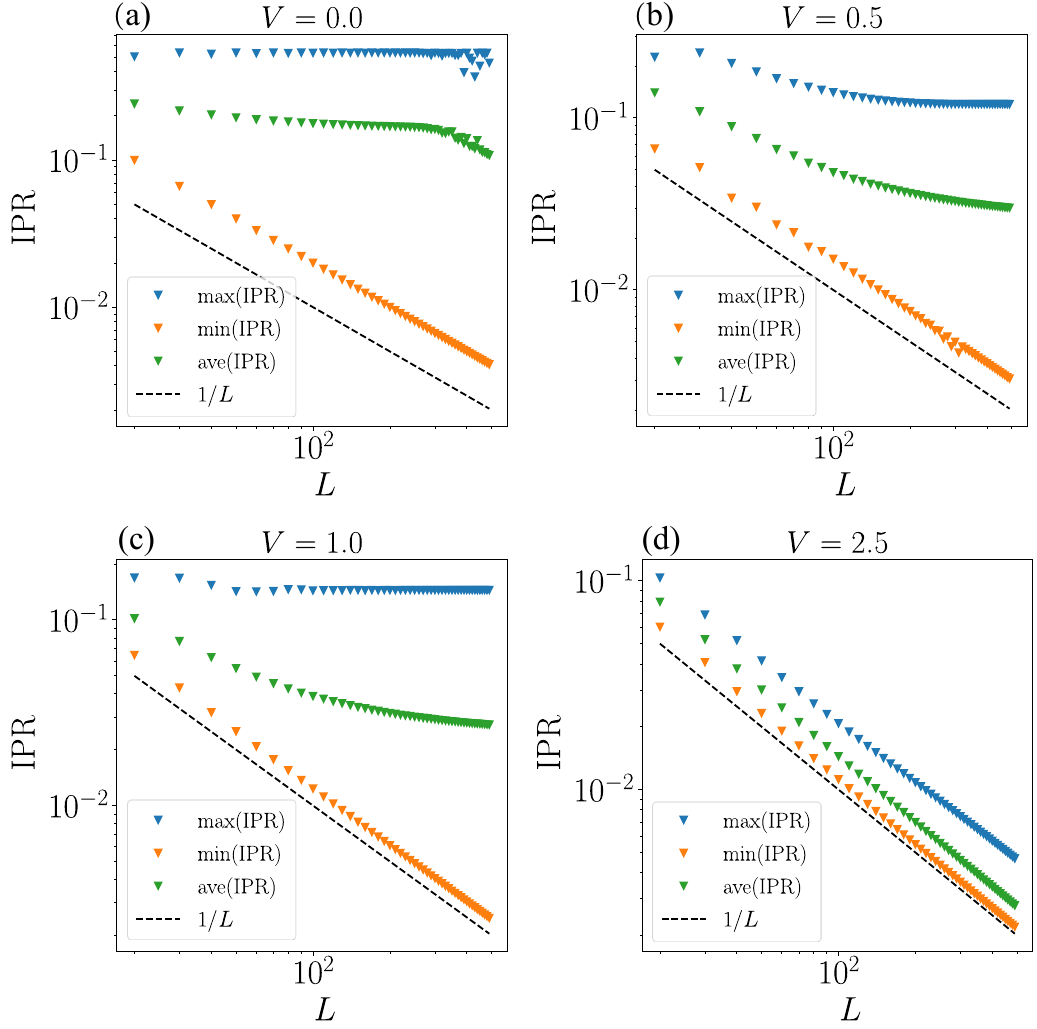}
\par\end{centering}
\caption{Scaling of the maximum, minimum and average of IPRs versus the length
of ladder $N$ under MBC. Other system parameters
are $(J,\gamma,\mu)=(1,0.3,0.5)$ for (a)--(d) and the length of
ladder goes from $N=10$ up to $N=250$ with $L=2N$.\label{fig:IPR_mbc}}
\end{figure}

Last but importantly, we investigate the scaling of IPRs versus the system size under OBC and MBC. To further distinguish between (skin) localized
and extended states in our model, we study the maximum, minimum and average of IPRs, which are defined as
\begin{alignat}{1}
\max({\rm IPR})& = \max_{i\in\{1,...,L\}}{\rm IPR}(|\psi_{i}\rangle),\label{eq:IPRmax}\\
\min({\rm IPR})& = \min_{i\in\{1,...,L\}}{\rm IPR}(|\psi_{i}\rangle),\label{eq:IPRmin}\\
{\rm ave}({\rm IPR})& = \frac{1}{L}\sum_{i=1}^{L}{\rm IPR}(|\psi_{i}\rangle),\label{eq:IPRave}
\end{alignat}
where $L=2N$ counts the total number of states. For extended (localized) states, we have ${\rm IPR}\sim N^{-1}$ (${\rm IPR}\sim\xi^{-1}$), where the localization length $\xi$ is independent of the system size. Our numerical results are presented for OBC in Fig.~\ref{fig:IPR_obc} and for MBC in Fig.~\ref{fig:IPR_mbc}. 

We find that under OBC, only the minimum of IPRs scales with the system size for some small interchain couplings, as shown in Fig.~\ref{fig:IPR_obc}(b)--(c), while the average and maximum of IPRs are almost constant. It also indicates that eventually the minimum of IPRs becomes constant for large $V$, as seen in Fig.~\ref{fig:IPR_obc}(d). This is in line with our previous discussions, i.e., our model effectively consists of two open HN chains at large $V$. Note that localization lengths of all eigenstates of an open HN chain are the same in the thermodynamic limit, while they may differ due to numerical calculations up to a finite system size. In short, we clearly observe the propagation of the non-Hermitian skin localization from the HN chain to another chain at its proximity, and all states can be localized in the hybridized ladder under OBC. 

We also observe the coexistence of finite amounts of extended and skin localized states in our model under MBC, though IPRs behave differently at large $V$ from those under OBC. We further numerically verify that the pseudo mobility edge phase under MBC agrees very well with the range of interchain coupling $V\in(V_{-}^{c},V_{+}^{c})$. This corresponds to the cases shown in Figs.~\ref{fig:IPR_mbc}(b)--(c), where the $\max({\rm IPR})\sim\xi^{-1}$ and the $\min({\rm IPR})\sim N^{-1}$. Now let us relate these results here to the previous results on spectrum under PBC. Remarkably, it turns out that the central point-gap phase 
under PBC corresponds to the pseudo mobility edge phase under MBC. When $|V|>|V_{\pm}^{c}|$, we find ${\rm IPR}\sim N^{-1}$ for all states, implying the coincidence between the line-gap phase within PBC and extended phase under MBC of the hybridized HN ladder. Because of the excellent phase correspondence observed here, we will focus on their intrinsic connections in the following subsections.

\subsection{Non-Hermitian skin localization transition and pseudo mobility edge\label{subsec:FD}}

\begin{figure}
\begin{centering}
\includegraphics[scale=0.49]{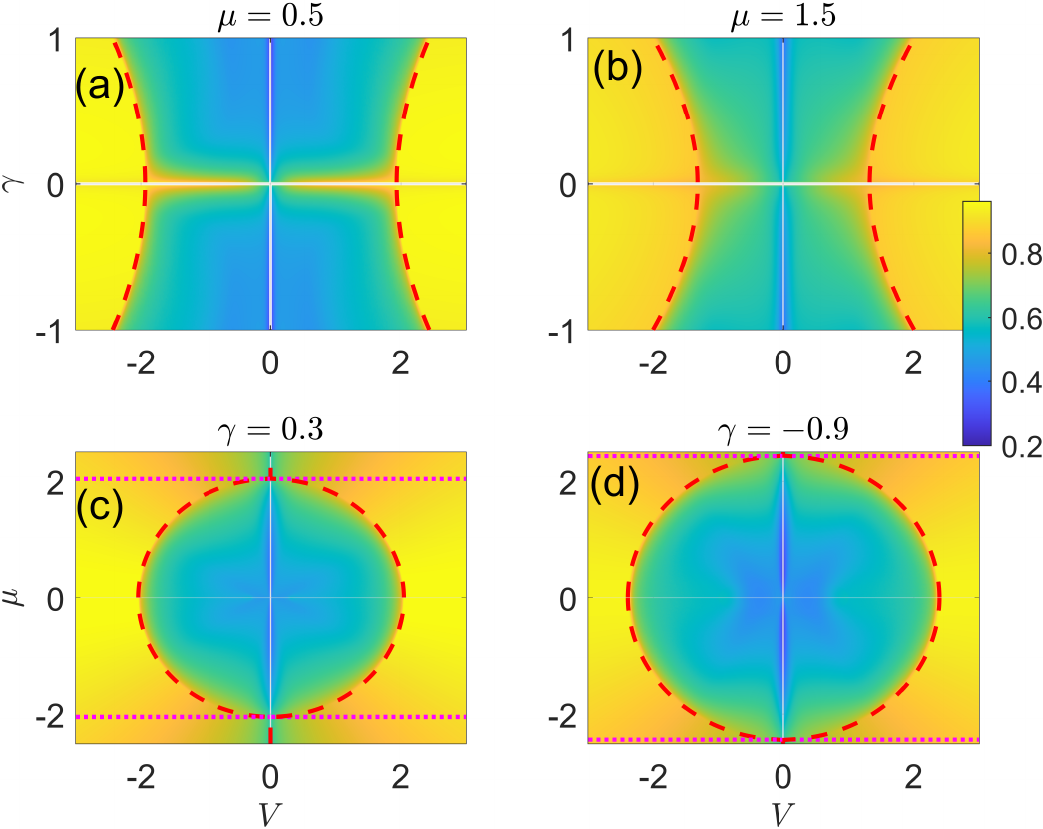}
\par\end{centering}
\caption{Fractal dimensional ${\rm FD}$ versus the hopping asymmetry $\gamma$
and interchain coupling $V$ (bias $\mu$) in (a)--(b) {[}(c)--(d){]}.
OBC/PBC is taken along leg A/B. The length of ladder is $N=500$. Red dashed and dotted lines are phase boundaries and
bounds of $\mu$ as given by Eqs.~(\ref{eq:Vc}) and (\ref{eq:MUpm}).\label{fig:FD}}
\end{figure}

We now know that tuning the interchain coupling in our model may induce a transition from a pseudo mobility edge phase to an extended phase under MBC. Surprisingly, we also have clues that this phase transition coincides with that from a point-gapped spectrum with winding number $w=\pm2$ to a line-gapped spectrum with $w=0$ under PBC. To obtain further confirmations of this view, we construct the FD of the eigenstates from the average of their IPRs, i.e.,
\begin{equation}
{\rm FD}=-\ln[{\rm ave}({\rm IPR})]/\ln L,\label{eq:FD}
\end{equation}
where $L=2N$ is the total number of states and ${\rm ave}({\rm IPR})$ is defined in Eq.~(\ref{eq:IPRave}). For a phase in which all states are extended, we would expect ${\rm ave}({\rm IPR})\sim L^{-1}$ and ${\rm FD}\simeq1$. For a phase with pseudo mobility edges, we would instead have $0<{\rm FD}<1$. In Fig.~\ref{fig:FD}, we show the FD of the hybridized HN ladder under MBC. In all these examples, we find that for $\gamma\neq0$, the configurations of FD exhibit two distinct regions. When $V\in(V_{-}^{c},V_{+}^{c})$, we find ${\rm FD}\in(0,1)$, which corresponds to the pseudo mobility edge phase in which extended and skin localized states coexist. When $|V|>|V_{\pm}^{c}|$, the FD approaches one. The system thus undergoes a delocalization transition and roams into an extended phase. Notably, the phase boundaries {[}dashed lines in Figs.~\ref{fig:FD}(a)--(d){]} are exactly determined by Eq.~(\ref{eq:Vc}), i.e., the transition point of the spectrum between point- and line-gapped under PBC. Moreover, beyond $\mu=\mu_{\pm}$ {[}dotted lines in Figs.~\ref{fig:FD}(c)--(d){]}, the system enters the extended phase whenever an infinitesimal coupling $V$ is switched on. In this case, the non-Hermitian skin localization of an open HN chain is vanishing by coupling it to a periodic chain, because of the finite boundary couplings inherited from its proximity. Hence the extra boundary coupling in the Hermitian chain is essential for the presence of delocalization transitions in our system under MBC.

\begin{figure}
	\begin{centering}
		\includegraphics[scale=0.49]{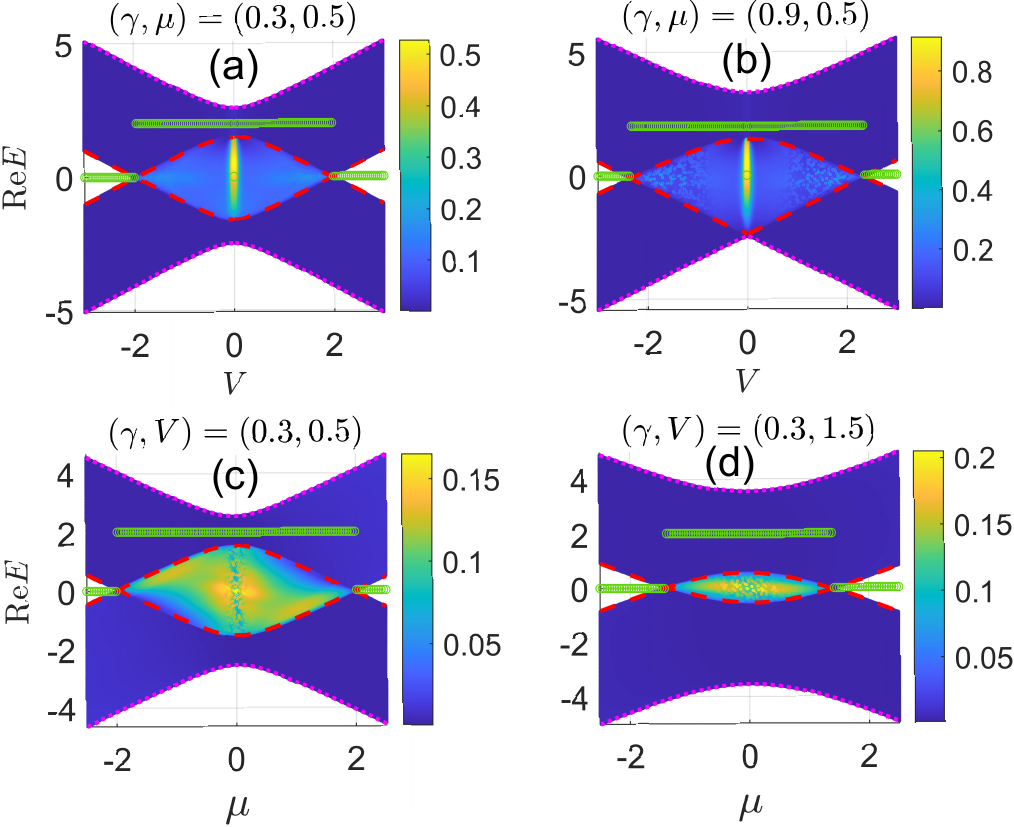}
		\par\end{centering}
	\caption{IPRs of all states versus the real parts of their energies ${\rm Re}E$ and interleg coupling $V$ (bias $\mu$) in (a)--(b) {[}(c)--(d){]}. The hybridized HN ladder contains $N=500$ unit cells. OBC/PBC along the leg A/B is taken in the calculation of IPRs. Dashed and dotted lines are bounds of the mobility edge (in ${\rm Re}E$) and the real parts of spectrum, as obtained from Eqs.~(\ref{eq:Ep0})--(\ref{eq:Emp}). Green circles denote the winding number $w$, generated under the PBC by Eq.~(\ref{eq:WN}).\label{fig:ME1}}
\end{figure}

To acquire more details of the pseudo mobility edge, we calculate the IPRs of all eigenstates with respect to the real parts of their energies. Our results are presented in Fig.~\ref{fig:ME1} with considerations of other parameters as well. The red
dashed lines in Figs.~\ref{fig:ME1}(a)--(d) are band edges determined from Eqs.~(\ref{eq:Em0}) and (\ref{eq:Epp}), beyond which the IPRs of all states go to zero in the $L\rightarrow\infty$ limit. Therefore,
Eqs.~(\ref{eq:Em0}) and (\ref{eq:Epp}) can be viewed as the upper and lower bounds of the pseudo mobility edge in the ${\rm Re}E$-parameter plane of the system for $V\in(V_{-}^{c},V_{+}^{c})$ and $\mu\in(\mu_{-},\mu_{+})$.
Moreover, we observe a quantized jump of the spectral winding number $w$ under PBC {[}green circles in Figs.~\ref{fig:ME1}(a)--(d){]} from two to zero when the system transits from the pseudo mobility edge phase to the extended phase under MBC. These results confirm that the spectral winding number $w$ defined in Eq.~(\ref{eq:WN}) can serve as a topological order parameter to characterize the phase transitions in the hybridized HN ladder under MBC. Our key results are collected in Table \ref{tab:Sum} for quick a reference. 

\subsection{Topological phase diagram\label{subsec:WN}}
\begin{figure}
\begin{centering}
\includegraphics[scale=0.49]{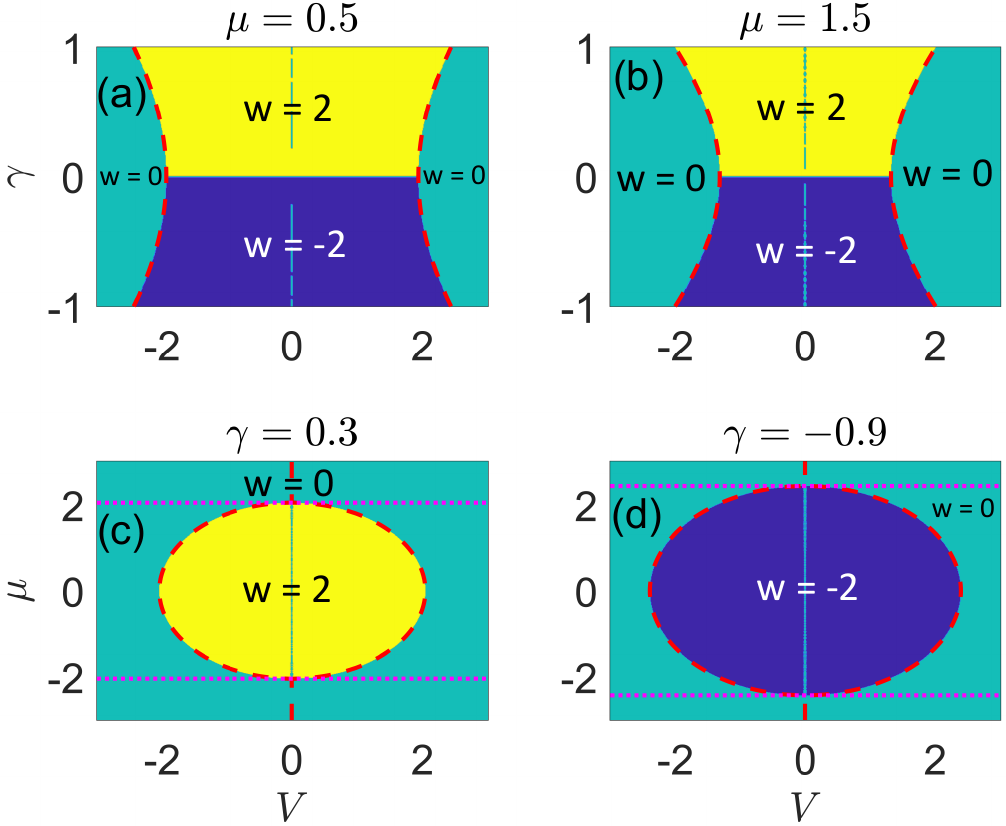}
\par\end{centering}
\caption{Phase diagram of the hybridized HN ladder. In (a)--(d), each region with a uniform color corresponds to a topological phase, whose winding number $w$ is denoted explicitly therein. The red dashed and dotted lines denote phase boundaries and upper/lower bounds of $\mu$ for the NHSE, which are obtained from Eqs.~(\ref{eq:Vc}) and (\ref{eq:MUpm}).\label{fig:WN}}
\end{figure}

Based on the above discussions, let us now complete the topological characterization of our system. In Fig.~\ref{fig:WN}, we present typical examples of topological phase diagrams of the hybridized HN ladder. We first note that the sign of winding number $w$ is determined by the sign of $\gamma$, or physically speaking the ``chirality'' of the NHSE. More precisely, when $\gamma>0$ ($\gamma<0$), bulk states are mainly localized at the left (right) boundary of the ladder, and we find $w=2$ ($-2$) for $V\in(V_{-}^{c},V_{+}^{c})$ {[}in between the red dashed lines in Figs.~\ref{fig:WN}(a)--(d){]}. When the interchain coupling $|V|>|V_{\pm}^{c}|$, we find $w=0$ for all cases. Comparing the results in Figs.~\ref{fig:FD} and \ref{fig:WN}, we find that the pseudo mobility edge (extended) phase coincides well with the regions in which $w=\pm2$ ($w=0$). Hence, we establish a one-to-one correspondence between the spectrum (point-gapped or line-gapped under PBC), states (with or without skin localized modes under MBC) and topological properties of the hybridized HN ladder.

Our interpretation of $w$ as a topological order parameter for the system under PBC and MBC can be further understood as follows. When the point-gap in the range $E\in({\rm Re}E_+(\pi),{\rm Re}E_-(0))$ is present under the PBC, we find all bulk states appearing under the MBC to be localized with real energies in that energy range, while other bulk states are extended with non-real energies. When the point-gap in the overlapping region of the two bands vanishes such that $w$ changes from $\pm2$ to $0$, the real-energy localized states and the pseudo mobility edge in the spectrum also vanish. In this sense, we interpret the region with pseudo mobility edge as a phase that is topologically protected by the point-gap winding number of the PBC Hamiltonian. Note that this type of spectral winding number does not require any symmetry protection, and its quantization has been shown to be robust to various perturbations~\cite{gong2018}.

We may also appreciate the excellent correspondence observed in our system under PBC and MBC from another perspective. If we regard the missing coupling between the two ends in the HN chain as a defect in our system under MBC, then our result may also be interpreted as a ``bulk-defect correspondence'' in this quasi-1D non-Hermitian system. Therefore, the topological invariant of bulk spectrum under PBC is closely related to the localization nature of states around such defect in the ladder under MBC.

\begin{figure}
	\begin{centering}
		\includegraphics[scale=0.49]{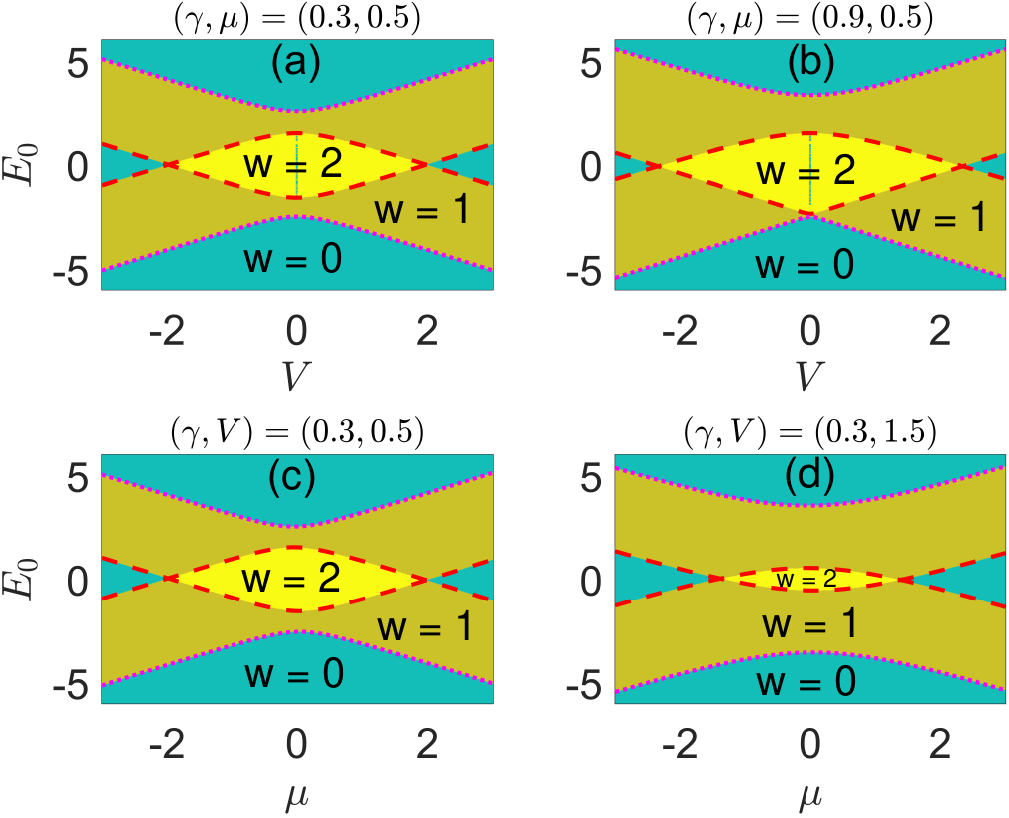}
		\par\end{centering}
	\caption{Winding number $w$ versus the base energy $E_{0}$ and interchain coupling $V$ (bias $\mu$) in (a)--(b) {[}(c)--(d){]}. In (a)--(d), each region with a uniform color has a fixed $w$. The dashed and dotted lines are band edges determined by Eqs.~(\ref{eq:Ep0})--(\ref{eq:Emp}).\label{fig:WNvsEB}}
\end{figure}

\begin{table*}
	\begin{centering}
		\begin{tabular}{|c|c|c|}
			\hline 
			Phase & Pseudo mobility edge & Extended\tabularnewline
			\hline 
			\hline 
			\multirow{2}{*}{Condition} & \multirow{2}{*}{$|\mu|<|\mu_{\pm}|,\,|V|<|V_{\pm}^{c}|$} & \multicolumn{1}{c|}{$|\mu|<|\mu_{\pm}|,\,|V|>|V_{\pm}^{c}|$}\tabularnewline
			&  & or $|\mu|\geq|\mu_{\pm}|,\,|V|>0$\tabularnewline
			\hline 
			\multirow{2}{*}{Energy} & Real/Complex for skin- & \multirow{2}{*}{Complex}\tabularnewline
			& localized/extended states & \tabularnewline
			\hline 
			\multirow{2}{*}{IPR} & $\sim\xi^{-1}$ / $\sim L^{-1}$ for skin- & \multirow{2}{*}{$\sim L^{-1}$}\tabularnewline
			& localized/extended states & \tabularnewline
			\hline 
			FD & $\in(0,1)$ & $\simeq1$\tabularnewline
			\hline 
			Winding & \multirow{2}{*}{$w=\pm2$} & \multirow{2}{*}{$w=0$}\tabularnewline
			number &  & \tabularnewline
			\hline 
		\end{tabular}
		\par\end{centering}
	\caption{Summary of results for the hybridized HN ladder under MBC. $V_{\pm}^{c}$
		and $\mu_{\pm}$ are given by Eqs.~(\ref{eq:Vc}) and (\ref{eq:MUpm}).
		$\xi$ and $L$ denote the localization length and total number
		of lattice sites. IPR, FD and $w$
		refer to the inverse participation ratio, fractal dimension~[see Eq.~(\ref{eq:FD})] and
		winding number~[see Eq.~(\ref{eq:WN})].\label{tab:Sum}}
\end{table*}

To furnish the topological view of the pseudo mobility edge under MBC, we let the base energy $E_{0}$ in Eq.~(\ref{eq:WN}) be a tunable parameter and show the winding number $w$ versus $(E_{0},V)$ and $(E_{0},\mu)$ in Fig.~\ref{fig:WNvsEB}. When $E_{0}$ is chosen beyond the range of bulk bands {[}i.e., $E_{0}<E_{-}(\pi)$ or $E_{0}>E_{+}(0)${]} or inside the line-gap under PBC, we find $w=0$ as expected. When $E_{0}$ is chosen inside the {central} point-gap 
{[}i.e., $E_{+}(\pi)<E_{0}<E_{-}(0)${]}, we find the winding number $w=2$ for $\gamma>0$, irrespective of the exact location of $E_{0}$ in this regime. 
This observation also indicates the robustness of the topological quantization of $w$ with respective to perturbations in system parameters $V$ and $\mu$.
Exceeding the bounds of the pseudo mobility edges {[}i.e., between the red dashed and dotted lines in Fig.~\ref{fig:WNvsEB}(a)--(d){]}, we find $w=1$ for all available $E_{0}$. Comparing these with the results presented in Fig.~\ref{fig:ME1}, we conclude that the winding number $w$ can indeed be employed as a refined facility to discriminate the states of the hybridized HN ladder into collections with distinct localization nature. For completeness, we have performed calculations of the spectrum, IPRs and FD by taking the other type of MBC, i.e., choosing PBC/OBC along the leg A/B of the ladder. Results thus obtained are consistent with those presented in this section, which confirms the generality of the connections we established among different aspects of the system. 

\section{Summary\label{sec:Sum}}

In this work, we considered a noninteracting coupled-chain system consisting of a non-Hermitian chain with nonreciprocal hoppings and a delocalized Hermitian chain. We discovered that a pseudo mobility edge phase may emerge under a weak interchain coupling. We demonstrated that non-Hermitian skin localization can propagate to its proximity when both chains are under OBC. Such NHSE propagation or takeover is fragile to extra boundary coupling in either of the two chains, reflecting the extreme sensitivity of non-Hermitian systems {and the fragility of the NHSE against boundary perturbations \cite{li2021impurity,Liu2021exact,Roccati2021,guo2021exact}}. We also characterized the transitions from the pseudo mobility edge phase to the extended phase under MBC with a topological spectral winding number under PBC of our model. Our findings thus provide more insights for understanding the topological and localization properties of the NHSE in a composite system, and unveil new possibilities of NHSE under external perturbations. In future work, it would be interesting to investigate the non-Hermitian localized system coupled to different systems~\cite{roccati2021exotic}, and consider possible realizations of the hybridized HN ladder in some quantum simulators. The potential applications of these results in sensing and lasing technologies are also worthy of further exploration. Another promising aspect may come from the consideration of interaction effects. Two Luttinger liquids coupled with single-particle hoppings have been intensively studied~\cite{Fabrizio1993,Clarke1994,Capponi1998,Luthra2008}. Effects brought by non-Hermiticity in one of the systems or both should be a fruitful problem to investigate. 

\begin{acknowledgments}
S.M. and L.Z. contributed equally to this work. L.Z. is supported by the National Natural Science Foundation of China (Grant No. 11905211). L.L. is supported by the National Natural Science Foundation of China (12104519) and the Guangdong Basic and Applied Basic Research Foundation (2020A1515110773), J.G. acknowledges support from Singapore National Research Foundation Grant No. NRF- NRFI2017-04 (WBS No. R-144-000-378-281).
\end{acknowledgments}

\bibliography{references.bib}


\end{document}